\documentclass[aps,physrev,reprint,amssymb,amsmath,onecolumn]{revtex4-2}
\usepackage{graphicx}
\usepackage{dcolumn}
\usepackage{bm}
\usepackage[mathscr]{euscript}

\begin{document}
\title{Excitron-Induced Pair Fluctuations Reveal Superconductivity in the Electron Gas}

\author{Yasutami Takada}
\email{Contact Author: takada@issp.u-tokyo.ac.jp}
\affiliation{The Institute for Solid State Physics, The University of Tokyo, 
Kashiwa, Chiba 277-8581, Japan}

\date{\today}    
\begin{abstract}
Understanding how superconductivity can emerge in dilute electronic systems remains 
a central challenge in condensed matter physics. By performing first-principles 
calculations of the electron self-energy $\Sigma({\bm k},i\omega_n)$ in the 
low-density three-dimensional electron gas, we identify a sharp divergence 
at $r_s \approx 8$ and $T \approx 10^{-4}\varepsilon_{\rm F}$, signaling a second-order 
phase transition. This critical behavior originates from one-dimensional superconducting 
fluctuations mediated by virtual excitations of an excitron---a quasi-1D 
electronic composite formed by an electron and longitudinal electron-hole pairs. 
Although the superconducting mechanism itself is plasmon-mediated, the excitron 
channel provides a unique window into its fluctuation dynamics. Near the transition, 
we observe a pseudogap and a linear-in-$T$ inverse electron lifetime, reminiscent of 
phenomena in high-$T_c$ materials. These results reveal an unexpected route 
by which plasmon-driven superconductivity manifests in the dilute 3D electron gas 
through quasi-1D excitron dynamics.
\end{abstract}
\maketitle
\section{Introduction}
\label{sec:1}

An accurate non-empirical treatment of the Coulomb interaction is essential 
for reliably calculating the superconducting transition temperature $T_c$ 
from first principles~\cite{YT_1980,Richardson_1997,Kurth_1999,Monthoux _2007,
Akashi_2013,Akashi_2014,YT_2015,Ruhman_2016,Wei_2021,Akashi_2022,Wang_2023,
Pellegrini_2023,Cai_2025}. This issue is closely tied to the long-standing 
question of whether the three-dimensional homogeneous electron gas (3DHEG), 
an assembly of $N$ electrons embedded in a 3D uniform positive rigid background, 
exhibits superconductivity without phonons. Previous studies have invoked 
either the Kohn-Luttinger (KL) mechanism~\cite{Kohn_1965} or 
the dynamic screening (plasmon) mechanism originally proposed by the present 
author~\cite{YT_1978,Rietschel_1983,Grabowski_1984,
Buche_1990,YT_1992a,YT_1992b,YT_1993,Richardson_1996,Kukkonen_2021}. 

The KL mechanism has recently been shown to be irrelevant for the 3DHEG~\cite{Cai_2022}, 
but it remains controversial whether dynamic screening can yield superconductivity 
with $T_c \! \gtrsim \! 1$~K for $r_s \! \lesssim \! 10$, where $r_s$ is the 
usual density parameter, related to the Fermi momentum $k_{\rm F}$ through 
$r_s\!=\! 1.919/k_F$ in units of the Bohr radius. (We use units with 
$\hbar\!=\!k_{\rm B}\!=\!1$.) Earlier works estimated $T_c$ 
by solving the gap equation using the Kukkonen-Overhauser (KO) 
ansatz~\cite{Kukkonen_1979} for the effective interaction. This approach 
proceeds from $T$ below $T_c$, and the resulting $T_c$ depends sensitively 
on the details of the KO interaction, preventing a definite conclusion. 

Because superconductivity is a second-order phase transition, 
$T_c$ can also be determined from the normal state by identifying singular 
behavior in fluctuation quantities. In particular, the Cooper-pair 
fluctuation propagator $D_{sc}(Q)$ diverges at $Q=0$ and $T\!=\!T_c$, 
where $Q \!\equiv\! \{{\bm q},i\omega_q\}$ denotes momentum 
and bosonic Matsubara frequency. A standard strategy is to compute $D_{sc}(Q)$ 
via the Bethe-Salpeter equation (BSE) using the KO interaction as the kernel, 
but this yields essentially the same $T_c$ as the gap-equation approach 
and thus does not resolve the controversy. 

In view of this situation, our strategy is twofold: (i) to obtain the 
self-energy $\Sigma(K)$ at various $r_s$ and $T$ in the normal phase 
without relying on the KO ansatz, and (ii) to search for signatures 
of a superconducting phase transition in $\Sigma(K)$. Here, $K\!\equiv \!
\{{\bm k},i\omega_n\}$ denotes momentum and fermionic Matsubara frequency. 
In the first step, we employ a recently-developed fully self-consistent scheme 
to determine $\Sigma(K)$ accurately~\cite{YT_2024}, and in the second, 
we analyze the resulting $\Sigma(K)$ by comparison with previous results 
obtained in the presence of superconducting fluctuations~\cite{Abrahams_1970,Hurault_1970,Takayama_1970}.

Implementing this strategy reveals a divergence in the renormalization function 
$Z(K)\! \equiv \! 1\!-\!{\rm Im}[\Sigma(K)]/\omega_n$, indicating a phase transition 
at $r_s\! \approx \! 8$ and $T \! \approx \! 10^{-4}\varepsilon_{\rm F}$ 
(with $\varepsilon_{\rm F}$ being the Fermi energy). 
This divergence arises from enhanced one-dimensional (1D) superconducting 
fluctuations near $T_c$, generated by virtual excitations of 
an excitron---a quasi-1D electronic composite propagating along the associated 
polarization field~\cite{YT_2024}. In the transition region, 
we also find a $T$-dependent dip (or a pseudogap) in the density of states 
near the Fermi level and a linear-in-$T$ inverse electron lifetime. 
These results demonstrate superconductivity in the 3DHEG at $T_c\! \approx \! 1$ K 
for $r_s \! \approx \! 8$, consistent with previous estimates 
in Refs.~\cite{YT_1992b,YT_1993}. 

\section{Functional GW$\mbox{\boldmath$\Gamma$}$ (FGW$\mbox{\boldmath$\Gamma$}$) Method 
for calculating $\mbox{\boldmath$\Sigma$}${\textbf ({\textit K})}}
\label{sec:2}

The key idea of our approach is that superconducting fluctuations need not be modeled 
explicitly. Instead, once the normal-state self-energy $\Sigma(K)$ is determined 
with controlled accuracy and without violating conservation laws, all fluctuation 
effects---including those associated with superconductivity---are automatically 
encoded in it. This motivates us to construct a fully conserving and self-consistent 
scheme for $\Sigma(K)$ in the 3DHEG.

The self-energy is rigorously given by
\begin{align}
\label{eq:01}
\Sigma(K)\! =\! -\sum_{Q} W(Q) G (K\!+\!Q)\Gamma(K,K\!+\!Q),
\end{align}
where $W(Q)\!=\! V({\bm q})/[1\!+\!V({\bm q})\Pi(Q)]$ employs the well-established 
polarization function $\Pi(Q)$ from analytic~\cite{Richardson_1994,YT_2016} 
and quantum Monte Carlo (QMC)~\cite{Moroni_1995,Dornheim_2018,Haule_2019,
LeBlanc_2022,Chuna_2025} studies. The Dyson equation 
\begin{align}
G^{-1}(K) = G_0^{-1}(K)- \Sigma(K)=Z(K)i\omega_n-E(K).
\label{eq:02}
\end{align} 
provides the standard decomposition of $\Sigma(K)$ into its real and imaginary 
parts, where $G_0(K)\!=\!(i\omega_n \!-\!\varepsilon_{{\bm k}})^{-1}$ with 
$\varepsilon_{{\bm k}}$ being the bare electron dispersion, given by 
$\varepsilon_{{\bm k}}\!=\!{\bm k}^2/(2m)\!-\!\varepsilon_{\rm F}$. Here, 
$m$ is the mass of a free electron.

With $W(Q)$ fixed, determining the vertex $\Gamma(K,K\!+\!Q)$ becomes central. 
In Hedin's formalism~\cite{Hedin_1965}, $\Gamma$ satisfies a BSE 
with kernel $\tilde{I}\!=\!\delta \Sigma/\delta G$, but practical 
choices such as GW or GW+BSE~\cite{Onida_2002} violate the Ward identity 
(WI)~\cite{Ward_1950}, leading to charge nonconservation and thus an 
unphysical self-energy. Since WI is essential for physically meaningful 
$\Sigma(K)$, these approximations are insufficient.

Enforcing WI at each iteration is known to greatly accelerate convergence 
toward the exact self-energy~\cite{YT_1995}, because WI tightly constrains the 
frequency and momentum dependence of $\Gamma$. 
Recent developments~\cite{YT_2001,Maebashi_2011,YT_2024} have produced a functional 
vertex $\Gamma(K,K\!+\!Q)$ (see Eq.~(SM5) in the Supplemental Material (SM)~\cite{SM}) 
satisfying WI together with standard conservation laws and sum rules. 
Using this vertex closes the self-consistent loop for $\Sigma(K)$, yielding 
a normal-state description in which fluctuation effects are naturally embedded.

We solve Eqs.~(\ref{eq:01}) and (\ref{eq:02}) self-consistently starting from the RPA 
self-energy. The resulting $\Sigma(K)$ is analytically continued to 
$\Sigma^R({\bm k},\omega)$ using Pad\'{e}~\cite{Vidberg_1977} and 
Nevanlinna~\cite{Fei_2021} schemes, which yield consistent low-energy results. 
We refer to this approach as the functional GW$\Gamma$ (FGW$\Gamma$) method.

As detailed in Ref.~\cite{YT_2024}, the FGW$\Gamma$ reproduces accurate quasiparticle (QP) 
renormalization factors $z^{*}$ and effective masses $m^{*}$ for metallic densities 
$1\!<\!r_s\!<\!6$, in excellent agreement with recent QMC simulations~\cite{Holzmann_2011,Hiraoka_2020,Haule_2022,Holzmann_2023} 
and experiments~\cite{Hiraoka_2020,Suortti_2000,Huotari_2010}. A weak low-energy 
peak (``excitron'') appears in the spectral function 
$A({\bm k},\omega) [\equiv \!-{\rm Im}\,G^{R}({\bm k},\omega)/\pi]$ at 
$T \lesssim 10^{-3}\varepsilon_{\rm F}$, but its small weight ensures 
that bulk properties remain governed by conventional quasiparticles.

\section{Calculated results of $\mbox{\boldmath$\Sigma$}${\textbf ({\textit K})} 
and related quantities}
\label{sec:3}

\begin{figure*}[tbhp]
\begin{center}
\includegraphics[width=1.0\linewidth,keepaspectratio,clip]{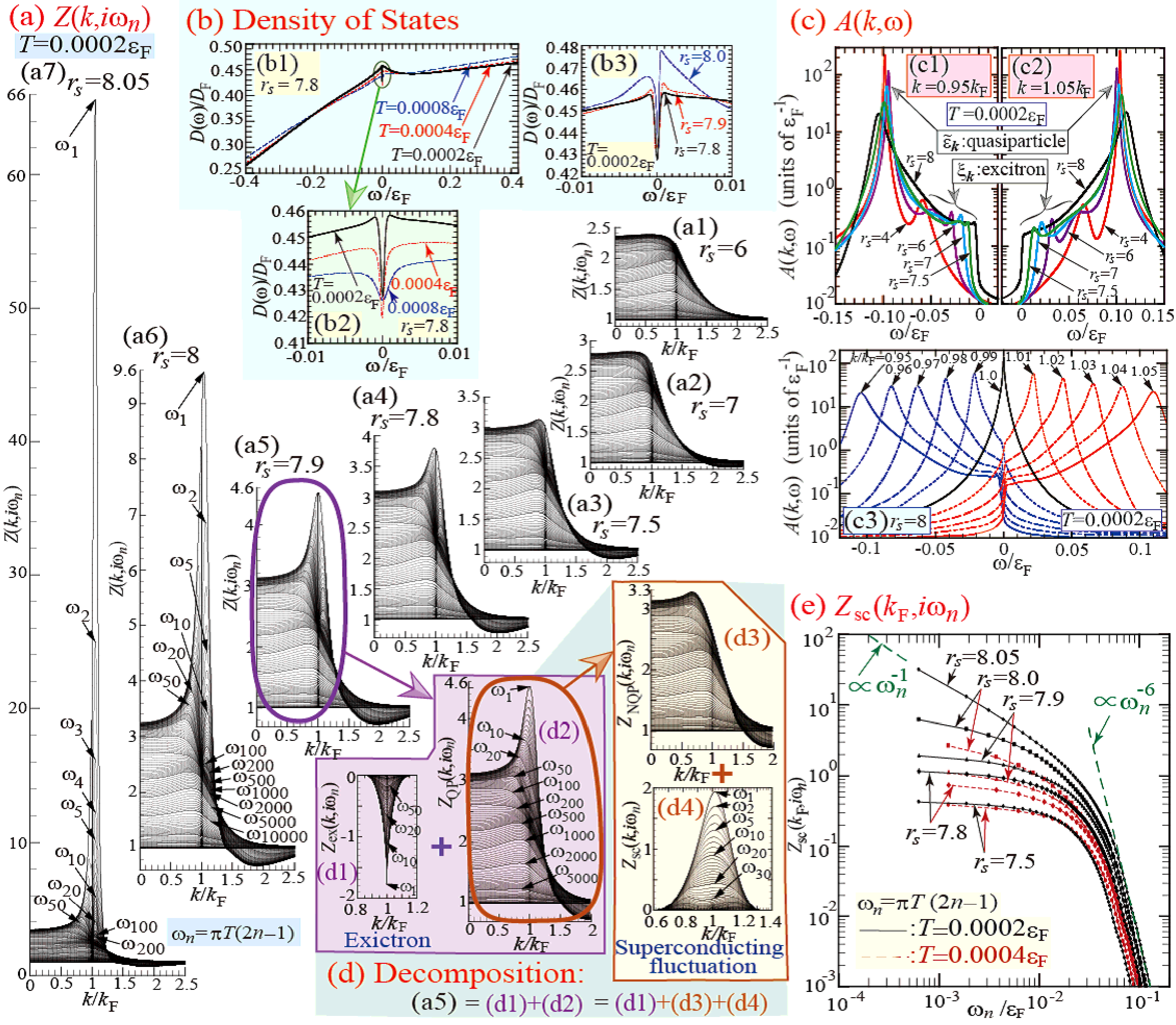}
\end{center}
\caption[Fig.01] {
(a) Renormalization function $Z(k,i\omega_n)$ at $T/\varepsilon_{\rm F}=0.0002$ 
for $r_s=6-8.05$ [panels (a1)-(a7)]. A sharp peak develops near $k\approx k_{\rm F}$ 
at $r_s\approx 8$, signaling a phase transition.
(b) Density of states $D(\omega)$ near the Fermi level: for $r_s=7.8$, 
$T/\varepsilon_{\rm F}=0.0002-0.0008$ [panels (b1),(b2)], and 
for $T/\varepsilon_{\rm F}=0.0002$, $r_s=7.8-8.0$ [panel (b3)]. A pseudogap 
appears, indicating superconducting-gap formation. 
(c) Spectral function $A(k,\omega)$ at $T/\varepsilon_{\rm F}=0.0002$ 
for $r_s=4-8$. Panels (c1),(c2) ($k/k_{\rm F}=0.95$, $1.05$) show that the QP dispersion 
$\tilde{\varepsilon}_{\bm k}$ is nearly $r_s$-independent, whereas the QP lifetime $\tau$, 
excitron dispersion $\xi_{\bm k}$, and excitron line shape vary strongly. 
At $r_s=8$ [panel (c3)], a broad QP peak and an excitron branch-cut structure 
with $\xi_{\bm k}/\tilde{\varepsilon}_{\bm k}\approx 0.02$ appear. 
(d) Procedure for extracting the superconducting-fluctuation component $Z_{\rm sc}(k,i\omega_n)$ 
[panel (d4)] from the full renormalization function $Z(k,i\omega_n)$ [panel (a5)].
(e) $Z_{\rm sc}(k_{\rm F},i\omega_n)$ for $r_s=7.5-8.05$ at $T/\varepsilon_{\rm F}=0.0002$ 
and for $r_s=7.5-8.0$ at $T/\varepsilon_{\rm F}=0.0004$. At large $\omega_n$, 
$Z_{\rm sc}(k_{\rm F},i\omega_n)\propto \omega_n^{-6}$; toward $\omega_n \to 0$, 
it increases as $\omega_n^{-\lambda}$ with $\lambda \approx 1$.
}
\label{fig:01}
\end{figure*}

We apply the FGW$\Gamma$ to the lower-density 3DHEG for $T\!=\!(2-8)\!\times\! 10^{-4}
\varepsilon_{\rm F}$. Figure~\ref{fig:01}(a) shows the evolution of $Z(K)$ 
with increasing $r_s$ from $6$ to $8.05$ at $T=0.0002\varepsilon_{\rm F}$. 
For $r_s \!<\!7.5$ [Figs.~\ref{fig:01}(a1),(a2)], $Z(K)$ exhibits the same sharp dip 
at the Fermi level as at $r_s=3.93$ in Ref.~\cite{YT_2024}, originating from 
the quasi-1D excitron. Its near-independence of $T$ indicates that the 3DHEG 
remains normal metallic up to $r_s\sim 7.5$ (see Fig.~SM1 in the SM~\cite{SM}). 
For $r_s \gtrsim 7.5$, however, a new bump emerges at the Fermi level in addition 
to the excitron dip.
 
As shown in Figs.~\ref{fig:01}(a3)-(a7), this bump rapidly develops into a sharp 
peak for $r_s\gtrsim 8$. No convergent $\Sigma(K)$ is obtained for $r_s\!\geq \!8.07$ 
at this $T$, even after thousands of iterations. 
A similar rapid change appears when varying $T$ at $r_s \gtrsim 7.8$ 
(see Fig.~SM2 in the SM~\cite{SM}). We attribute this bump-to-peak crossover 
to fluctuation phenomena characteristic of a second-order phase transition 
triggered by small changes in $r_s$ or $T$.

Aside from superconductivity~\cite{YT_1992b,YT_1993}, no second-order phase transition 
has been proposed in the 3DHEG in this density and temperature range. We therefore 
interpret this anomaly in $\Sigma(K)$ as arising from superconducting fluctuations 
(SCF). Below, we examine several quantities supporting this interpretation.

Unlike $Z(K)$, the real part $E(K)$ exhibits no significant anomaly (see Fig.~SM3 
in the SM~\cite{SM}). However, combining $E(K)$ and $Z(K)$ enables us to compute 
the spin-summed interacting density of states, $D(\omega)$, as detailed in Appendix A: 
\begin{align}
D(\omega)=-\frac{1}{\pi} {\rm Im}\sum_{{\bm k}\sigma}G^{R}({\bm k},\omega)
=2\sum_{\bm k}A({\bm k},\omega).
\label{eq:03}
\end{align}
Figure~\ref{fig:01}(b1) shows $D(\omega)$ near the Fermi level 
for temperatures ranging from $2\!\times\! 10^{-4}\varepsilon_{\rm F}$ to 
$8\!\times\! 10^{-4}\varepsilon_{\rm F}$ at $r_s\!=\!7.8$. Here, $D(\omega)$ 
is normalized by the noninteracting density of states at the Fermi level, 
$D_F=mk_{\rm F}/\pi^2$. By magnifying the horizontal axis by a factor of 40 
in Fig.~\ref{fig:01}(b2), we clearly observe a 
pseudogap---a $T$-dependent dip---in the immediate vicinity of 
the Fermi level. This feature becomes more pronounced as either $T$ 
or $r_s$ increases [Fig.~\ref{fig:01}(b3)], consistent with 
the interpretation in terms of SCF.

To ensure that Cooper-pairing-based SCF is compatible with a quasiparticle 
description, we examined $A({\bm k},\omega)$ from $r_s\!=\!4$ (sodium density) 
to $8$ at $T\!=0.0002\varepsilon_{\rm F}$. As shown in Fig.~\ref{fig:01}(c1),(c2), 
no qualitative change occurs, confirming that the QP picture remains valid 
throughout the transition region.

Quantitatively, four trends are important: (1) The QP dispersion 
$\tilde{\varepsilon}_{\bm k}$ is nearly independent of $r_s$, 
so $m^* \!\approx \!m$. 
(2) As $r_s \! \to \! 8$, the inverse QP lifetime $\tau^{-1}$ increases 
strongly, violating the usual $T^2$ behavior. 
(3) The excitron spectral shape evolves from Lorentzian at $r_s \! = \! 4$ 
to a branch cut singularity at $r_s \! \approx \! 8$, indicating 
an increasingly 1D nature. 
(4) The excitron dispersion $\xi_{\bm k}$ softens markedly, 
with $\xi_{\bm k}/\tilde{\varepsilon}_{\bm k}\! \to \! 0.02$ 
at $r_s\!=\! 8$ [Fig.~\ref{fig:01}(c3)], making $\xi_{\bm k}$ 
comparable to $T$ near the Fermi level. Thus, a macroscopic number 
of excitrons are thermally excited and act as QP scattering centers.

To analyze the anomaly in $Z(K)$, we first extract the excitron contribution 
$Z_{\rm ex}(K)$ using the smoothing procedure of Ref.~\cite{YT_2024}, 
yielding the QP part $Z_{\rm QP}(K)$. Applying the same procedure 
to $Z_{\rm QP}(K)$ isolates the anomalous peak $Z_{\rm sc}(K)$. 
This yields the decomposition 
$Z(K)\!=\!Z_{\rm NQP}(K)\!+\!Z_{\rm ex}(K)\!+\!Z_{\rm sc}(K)$, 
where $Z_{\rm NQP}(K)$ is the normal-state QP renormalization function 
defined within conventional Fermi-liquid theory. 
The extraction process at $r_s\!=\!7.9$ is illustrated in Fig.~1(d) 
(see also Fig.~SM4 in the SM~\cite{SM}).

Performing this for $r_s\! \geq \! 7.5$ yields $Z_{\rm sc}(K)$. 
Figure~\ref{fig:01}(e) shows $Z_{\rm sc}(k_{\rm F},i\omega_n)$ 
versus $\omega_n$ on a log-log scale. As $T$ decreases, 
$Z_{\rm sc}(k_{\rm F},i\omega_n)$ is enhanced for all $r_s$, 
especially at larger $r_s$. For $\omega_n \! \gtrsim \! 0.05 \varepsilon_{\rm F}$, 
$Z_{\rm sc}(k_{\rm F},i\omega_n)\! \propto \! \omega_n^{-6}$ 
regardless of $r_s$ or $T$. 
For $r_s \! \gtrsim \! 7.8$, $Z_{\rm sc}(k_{\rm F},i\omega_n)$ diverges 
as $\omega_n \! \to \! 0$, approaching $\omega_n^{-1}$ for $r_s\! \gtrsim \!8$. 
Although incompatible with conventional Fermi liquid theory, 
this behavior is consistent with dirty-superconductor 
theory~\cite{Abrahams_1970,Hurault_1970,Takayama_1970}, as discussed below.


\section{Analysis of the anomaly in $\mbox{\boldmath$\Sigma$}${\textbf ({\textit K})}}
\label{sec:4}

\begin{figure*}[tbhp]
\begin{center}
\includegraphics[scale=0.4,keepaspectratio,clip]{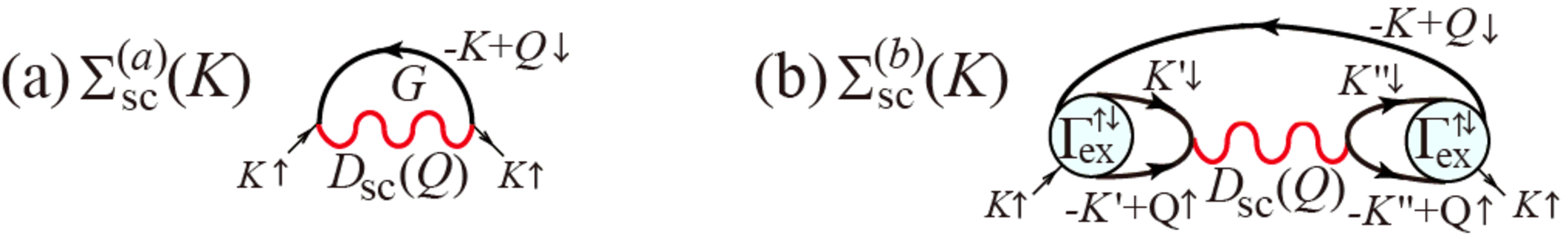}
\end{center}
\caption[Fig.02] {
Self-energy diagrams for up-spin electrons arising from 
(a) 3D superconducting fluctuations and (b) 1D ones in the presence of an excitron. 
Solid lines denote $G$, wavy lines $D_{\rm sc}$, and 
$\Gamma_{\rm ex}^{\uparrow \downarrow}$ is the excitron forming interaction vertex.  
}
\label{fig:02}
\end{figure*}

The lowest-order fluctuation contribution from $D_{\rm sc}(Q)$, 
shown in Fig.~\ref{fig:02}(a) as $\Sigma_{\rm sc}^{(a)}(K)$, is analytically evaluated 
near $T_c$ in a dirty superconductor (Appendix B). For large $\omega_n$, 
the resulting renormalization function $Z_{\rm sc}^{(a)}(k_{\rm F},i\omega_n)$ 
[$=-{\rm Im}\,\Sigma_{\rm sc}^{(a)}(K)/\omega_n$], decreases as $\omega_n^{-3/2}$ 
[Eq.~(\ref{eq:A8})], which is fundamentally incompatible with the $\omega_n^{-6}$ 
decay observed in $Z_{\rm sc}(k_{\rm F},i\omega_n)$ in Fig.~\ref{fig:01}(e). 
Thus $\Sigma_{\rm sc}^{(a)}(K)$ cannot account for the anomaly in $\Sigma(K)$.

A qualitatively new fluctuation channel emerges in the presence of an excitron, 
shown in Fig.~\ref{fig:02}(b) as $\Sigma_{\rm sc}^{(b)}(K)$. 
An excitron---an electron bound to a longitudinal polarization field---enables an incoming 
electron to form a transient Cooper pair with an opposite-spin electron from the field. 
This pair fluctuation propagates strictly along the polarization field and then dissociates, 
producing superconducting fluctuations that are effectively one dimensional. 
This 1D character is the key ingredient missing from conventional theories.

Appendix C provides an analytic evaluation of $\Sigma_{\rm sc}^{(b)}(K)$ near $T_c$. 
Using $A_{1}(k_{\rm F},i\omega_n)$ from Eq.~(\ref{eq:B5}), the corresponding 
renormalization function is
\begin{align}
\label{eq:04}
Z_{\rm sc}^{(b)}(k_{\rm F},i\omega_n) = 
\frac{1}{2} \left ( 1+\frac{1}{2\tau|\omega_n|} \right )
A_{1}(k_{\rm F},i\omega_n).
\end{align}
For large $\omega_n$, we obtain $Z_{\rm sc}^{(b)}(k_{\rm F},i\omega_n)
\!\propto\! \xi(T)\,\omega_n^{-6}$ [Eq.~(\ref{eq:B6})], with 
$\xi(T)\!=\!\sqrt{\pi D/8(T\!-\!T_c)}$. Here, $\xi(T)$ is the coherence length and 
$D$ is the QP diffusion contant. This reproduces the key features of 
Fig.~\ref{fig:01}(e): the rapid $\omega_n^{-6}$ decay and the enhancement of 
$Z_{\rm sc}(k_{\rm F},i\omega_n)$ upon lowering $T$. Thus $\Sigma_{\rm sc}^{(b)}(K)$ 
provides a natural and internally consistent explanation of the anomaly.

\begin{figure*}[bhtp]
\begin{center}
\includegraphics[width=0.9\linewidth,keepaspectratio,clip]{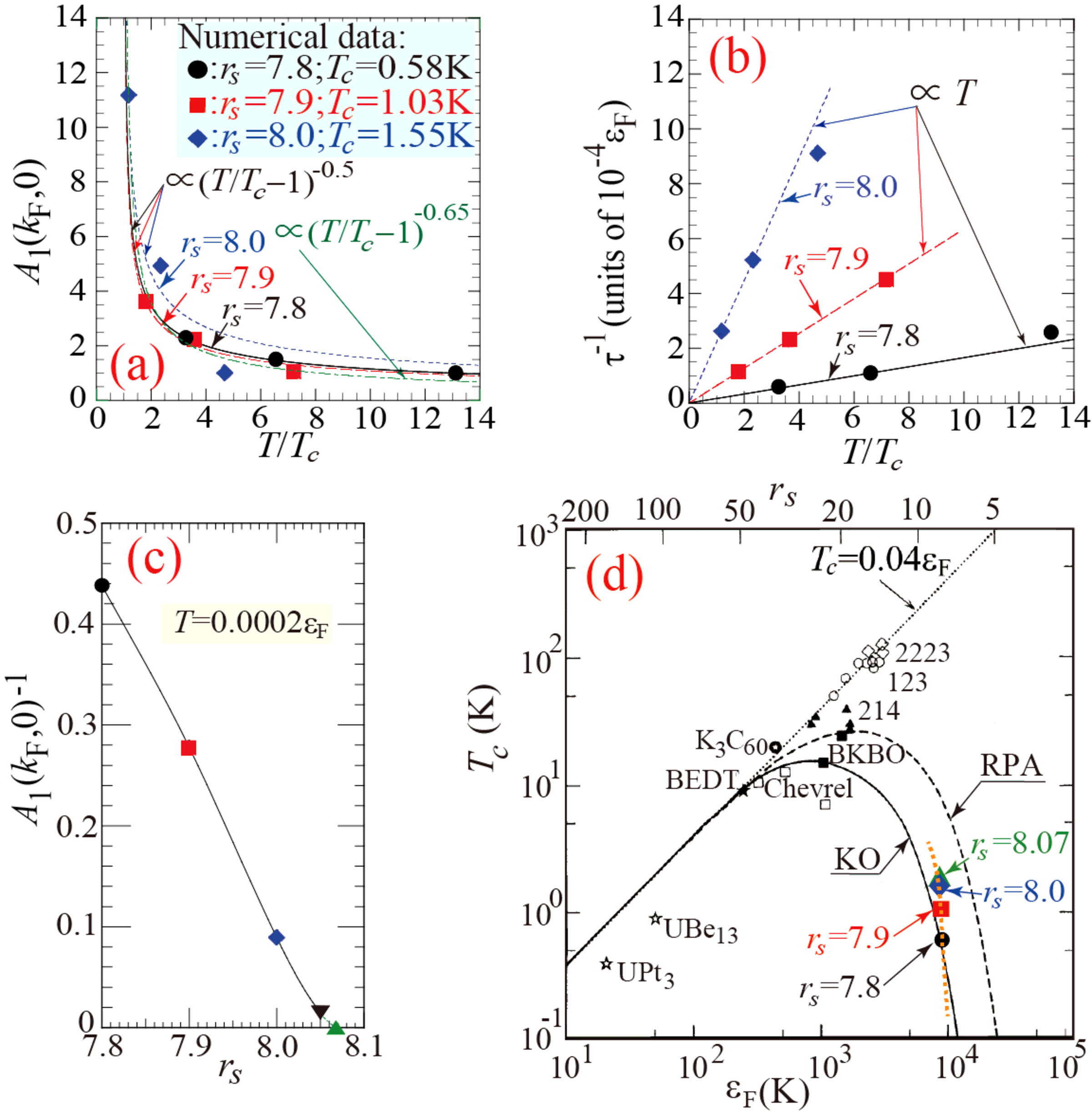}\end{center}
\caption[Fig.03]{
(a) Strength of the 1D superconducting-fluctuation contribution at $\omega_n\!=\!0$, 
$A_{1}(k_{\rm F},0)$, versus $T/T_c$ for $r_s\!=\!7.8-8.0$ with the corresponding 
$T_c$ values indicated. The scaling $A_{1}(k_{\rm F},0)\propto \xi(T)$ is reasonably 
satisfied. The coherence length follows $\xi(T) \propto (T/T_c\!-1\!)^{-\nu}$, 
with $\nu=0.5$ for $r_s=7.8$ and $7.9$, and $\nu=0.65\pm0.05$ for $r_s=8.0$. 
(b) Inverse QP lifetime $\tau^{-1}$ versus $T/T_c$ for $r_s\!=\!7.8-8.0$. 
A linear $T$-dependence persists over a wide range. 
(c) $A_{1}(k_{\rm F},0)^{-1}$ versus $r_s$ at $T/\varepsilon_{\rm F}=0.0002$. 
Extrapolation gives $A_{1}(k_{\rm F},0)^{-1}\to 0$ at $r_s \approx 8.07$, 
indicating the superconducting transition.
(d) Four $T_c$ values obtained at different $r_s$ in this study are plotted 
in Fig.~2 of Ref.~\cite{YT_1992b}. The orange dotted curve serves as a guide 
to the eye and shows reasonable agreement with the black solid curve obtained 
by solving the gap equation within the Kukkonen-Overhauser (KO) ansatz.
}
\label{fig:03}
\end{figure*}

Under the assumptions in Appendix D, we determine the limiting value $A_{1}(k_{\rm F},0)$ 
in Eq.~(\ref{eq:04}), together with $\tau$, for $r_s\!=\!7.8$, $7.9$, and $8.0$ 
at $T/\varepsilon_{\rm F}\!=\!0.0002$, $0.0004$, and $0.0008$. 
Figure~\ref{fig:03}(a) shows that $A_1(k_{\rm F},0)\! \propto \! \xi(T)
\! \propto \! \epsilon^{-\nu}$. The exponent $\nu$ takes the classical value $\nu\!=\!0.5$ 
at $r_s\!=\!7.8$ and $7.9$ for $T_c\!=\!0.58$K and $1.03$K, respectively.
At $r_s\!=\!8.0$, a similar trend is observed for $T_c\!=\!1.55$K, although a slightly 
better fit is achieved with $\nu\!=\!0.65 \pm 0.05$, consistent with critical fluctuations~\cite{Kiometzis_1994,Herbut_1996}.

Figure~\ref{fig:03}(b) shows that the extracted $\tau^{-1}$ scales linearly with $T$ 
over a wide range. This behavior arises naturally from scattering 
by thermally excited excitrons, whose density is proportional to $T$ 
due to their one-dimensional dispersion.

We also estimate $T_c$ by locating the value of $r_s$ at which 
$A_1(k_{\rm F},0)^{-1}\! \to \!0$ at a fixed $T/\varepsilon_{\rm F}$. 
Figure~\ref{fig:03}(c) yields $T_c=1.79$K at $r_s=8.07$. 

The four values of $T_c$ obtained here are plotted in Fig.~2 of Ref.~\cite{YT_1992b} 
[see Fig.~\ref{fig:03}(d)], showing reasonable agreement with the plasmon-mediated 
$T_c$ values obtained by solving the gap equation below $T_c$ 
using $\tilde{J}$ within the Kukkonen-Overhauser ansatz. Upon closer inspection, 
however, a systematic difference in slope appears near $r_s \sim 8$ 
between the orange dotted and black solid curves, suggesting that a refinement 
of the KO effective interaction is required to reduce this discrepancy.

\section{Conclusion and Discussion}
\label{sec:5}

In conclusion, first-principles calculations of $\Sigma(K)$ in the low-density 3DHEG 
reveal a sharp divergence at $r_s \approx 8$ and $T \approx 10^{-4}\varepsilon_{\rm F}$, 
marking a second-order phase transition. This critical behavior originates from 
one-dimensional superconducting fluctuations that manifest through virtual excitations 
of an excitron---a quasi-1D electronic composite formed by an electron and longitudinal 
electron–hole pairs. While superconductivity itself is driven by plasmon-mediated 
pairing as proposed in Refs.~\cite{YT_1978,YT_1992b,YT_1993}, the excitron channel 
provides a distinct window into its fluctuation dynamics. Near the transition, 
we find a pseudogap and a linear-in-$T$ inverse electron lifetime, revealing an unexpected 
route to superconductivity in a dilute electronic environment.


Three remarks are in order: 
(1) Ref.~\cite{YT_1993} reports similar $T_c$ values for $s$- and $p$-wave pairings 
near $r_s \approx 8$, but paramagnetic impurity effects were not included. 
Their inclusion leaves $s$-wave pairing intact (Anderson's theorem~\cite{Anderson_1959}), 
while suppressing $p$-wave superconductivity, implying that the superconductivity 
found here must be of $s$-wave character.
(2) As demonstrated in this work, the concept of the excitron establishes 
a new paradigm in many-body physics; it reveals a previously unrecognized 
fluctuation channel that governs superconducting critical behavior and offers 
a unified framework for the accompanying non-Fermi-liquid, or ``strange metal'', 
anomalies. 
(3) Extending excitron physics to the 2DHEG is an important next step. In particular, 
for superconductivity, it will be intriguing to examine how fluctuation effects 
manifest near the Berezinski\u{i}-Kosterlitz-Thouless 
transition~\cite{Berezinskii_1971,Kosterlitz_1973}.

\acknowledgments
The author thanks Hiroyuki Yata at the Institute for 
Solid State Physics, the University of Tokyo, for maintaining the cluster 
machines on which most of the computations were performed. 


\appendix

\setcounter{equation}{0}
\renewcommand{\theequation}{A\arabic{equation}}
\setcounter{figure}{0}
\renewcommand{\thefigure}{A\arabic{figure}}
\section{Evaluation of the density of states}

The density of states summed over spins, $D(\omega)$, in the interacting 
system is calculated from Eq.~(\ref{eq:03}) using the one-particle retarded 
Green's function $G^{R}({\bm k},\omega)$. 
For numerical evaluations, it is more effective to first calculate a quantity 
$\delta D(i\omega_n)$, defined by
\begin{align}
\delta D(i\omega_n)=-\frac{1}{\pi}\  \sum_{{\bm k}\sigma}\,
[G({\bm k},i\omega_n)-G_0({\bm k},i\omega_n)],
\label{eq:07d}
\end{align} 
and subsequently obtain $\delta D^{R}(\omega)$ by analytically continuing the result 
via Pad\`{e} approximants. Then, $D(\omega)$ is finally given by
\begin{align}
D(\omega)=D_0^R(\omega)+{\rm Im}\left [\delta D^{R}(\omega)\right ],
\label{eq:07e}
\end{align} 
with 
\begin{align}
D_0^R(\omega)=-\frac{1}{\pi}\  \sum_{{\bm k}\sigma}\,{\rm Im}\,G_0^R({\bm k},\omega)
=D_{\rm F}\sqrt{1+\frac{\omega}{\varepsilon_{\rm F}}}
\  \theta(\omega+\varepsilon_{\rm F}).
\label{eq:07f}
\end{align}

\setcounter{equation}{0}
\renewcommand{\theequation}{B\arabic{equation}}
\setcounter{figure}{0}
\renewcommand{\thefigure}{B\arabic{figure}}
\section{Effect of 3D Superconducting fluctuations on the self-energy}

\begin{figure*}[bthp]
\begin{center}
\includegraphics[scale=0.43,keepaspectratio,clip]{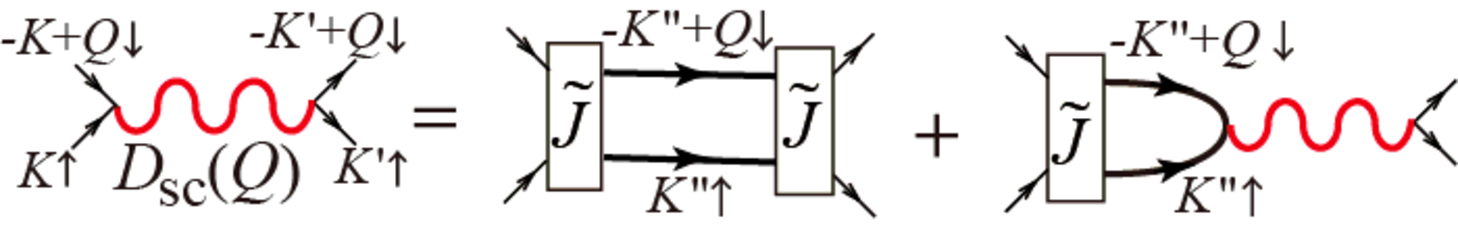}
\end{center}
\caption[Fig.04] {
Bethe-Salpeter equation for the Cooper-pair fluctuation 
propagator $D_{\rm sc}(Q)$, where $\tilde{J}$ denotes the electron-electron 
irreducible interaction.
}
\label{fig:04}
\end{figure*}

The lowest-order contribution from the Cooper-pair 
fluctuation propagator $D_{\rm sc}(Q)$ to $\Sigma(K)$, shown in Fig.~\ref{fig:02}(a) 
as $\Sigma_{\rm sc}^{(a)}(K)$, is 
\begin{align}
\label{eq:A1}
\Sigma_{\rm sc}^{(a)}(K)=-\sum_Q G(-K\!+\!Q)D_{\rm sc}(Q).
\end{align}
As diagrammatically shown in Fig.~\ref{fig:04}, $D_{\rm sc}(Q)$ and 
the corresponding $T_c$ are determined from the BSE 
with the electron-electron irreducible interaction $\tilde{J}$.
 
Near $T_c$, and for small $|{\bm q}|$ and $|\omega_q|$, 
irrespective of the details of $\tilde{J}$, $D_{\rm sc}(Q)$ is well approximated by
\begin{align}
\label{eq:A2}
D_{\rm sc}(Q)\approx \frac{1}{N_0}\ \frac{1}{\epsilon+A{\bm q}^2+B|\omega_q|},
\end{align}
with $\epsilon=(T-T_c)/T_c$ and $N_0$ being the QP density of states per spin 
at the Fermi level. For a dirty superconductor, $A=\pi D/(8T_c)$ and 
$B=\pi/(8T_c)$, where $D$ is the QP diffusion constant. 

Following Refs.~\cite{Abrahams_1970,Hurault_1970,Takayama_1970}, 
we evaluate $\Sigma_{\rm sc}^{(a)}(K)$ using the impurity-renormalized 
Green's function 
\begin{align}
\label{eq:A3}
G(K)  \approx   \frac{1}{i\tilde{\omega}_n\!-\!\varepsilon_{\bm k}},
\ {\rm with} \ 
\tilde{\omega}_n\!\equiv \!
\left ( 1\!+\!\frac{1}{2\tau |\omega_n|} \right )\omega_n,
\end{align}
where $\tau$ is the QP lifetime. Including vertex corrections 
due to impurity scatterings, the result at $k\!=\!k_{\rm F}$ is
\begin{align}
\label{eq:A4}
\Sigma_{\rm sc}^{(a)}(k_{\rm F},i\omega_n)=-\frac{1}{2}i\tilde{\omega}_n 
A_{3}(k_{\rm F},i\omega_n),
\end{align}
with 
\begin{align}
\label{eq:A5}
A_{3}(k_{\rm F},i\omega_n)=
\frac{T_c}{N_0AD^2}\sum_{{\bm q}}\frac{1}{{\bm q}^2+\epsilon/A}
\frac{1}{[{\bm q}^2+\epsilon/(2A)+|\omega_n|/D]^2}.
\end{align}
In deriving Eq.~(\ref{eq:A5}), we assume $\omega_n\!>\! \epsilon/B$. 
This condition holds for all Matsubara frequencies used in Fig.~\ref{fig:01}(e). 
Because the fluctuations propagate in 3D, the ${\bm q}$-integral is 
a three-dimensional one and can be evaluated analytically, yielding
\begin{align}
\label{eq:A6}
A_{3}(k_{\rm F},i\omega_n)= \frac{1}{N_0}
\left ( \frac{T_c}{\pi} \right )^2
\left [ \frac{\xi(T)}{D} \right ]^3
\frac{1}{a(a+1)^2},
\end{align}
where $\xi(T)$ is the coherence length, given by $\xi(T)\!=\!\sqrt{A/\epsilon}$, 
and the positive parameter $a$ is defined by the relation $a^2
=1/2+B|\omega_n|/\epsilon=1/2+|\omega_n|\xi(T)^2/D$.

For large $\omega_n$ ($\omega_n\! \gg \!\epsilon/B$), 
$a \!\approx \!\sqrt{|\omega_n|/D}\ \xi(T)\! \gg \!1$, giving  
\begin{align}
\label{eq:A7}
A_{3}(k_{\rm F},i\omega_n)\approx & \frac{1}{N_0}
\left ( \frac{T_c}{\pi} \right )^2
\left ( \frac{1}{D|\omega_n|} \right )^{3/2}.
\end{align}
Thus $Z_{\rm sc}^{(a)}(k_{\rm F},i\omega_n)$, associated with 
$\Sigma_{\rm sc}^{(a)}(K)$ through 
$Z_{\rm sc}^{(a)}(K)=-{\rm Im}\ \Sigma_{\rm sc}^{(a)}(K)/\omega_n$, behaves as
\begin{align}
\label{eq:A8}
Z_{\rm sc}^{(a)}(k_{\rm F},i\omega_n) 
\approx \ (1/2N_0)(T_c/\pi)^2 (D\omega_n)^{-3/2}.
\end{align}

\setcounter{equation}{0}
\renewcommand{\theequation}{C\arabic{equation}}
\setcounter{figure}{0}
\renewcommand{\thefigure}{C\arabic{figure}}
\section{Effect of excitron-induced 1D superconducting fluctuations on the self-energy}

\begin{figure*}[bthp]
\begin{center}
\includegraphics[scale=0.35,keepaspectratio,clip]{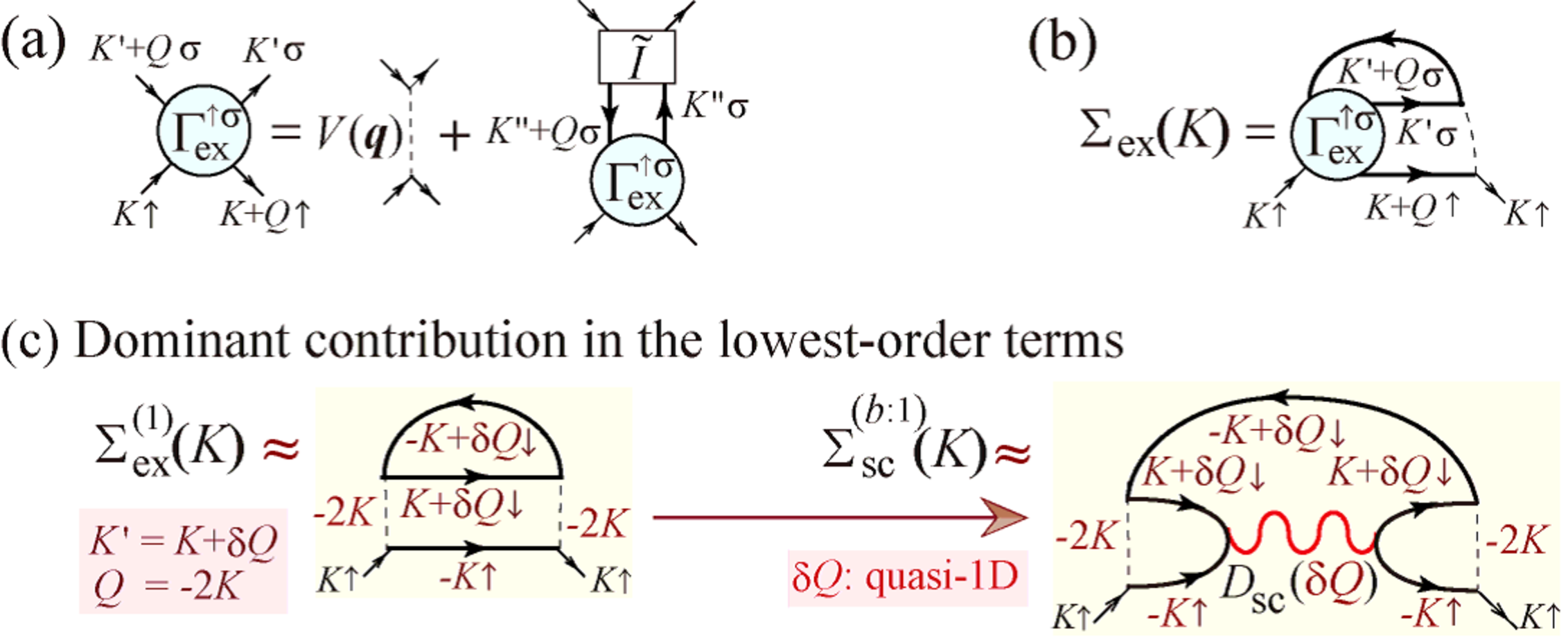}
\end{center}
\caption[Fig.05] {
(a) Bethe-Salpeter equation for the excitron-forming interaction vertex 
$\Gamma_{\rm ex}^{\uparrow \sigma}$, where $V({\bm q})$ is 
the bare Coulomb interaction and $\tilde{I}$ the electron-hole irreducible interaction. 
(b) Excitron contribution to the up-spin electron self-energy, $\Sigma_{\rm ex}$.
(c) Lowest-order dominant contribution to excitron formation, $\Sigma_{\rm ex}^{(1)}$, 
and the associated excitron-induced superconducting-fluctuation term 
in the up-spin electron self-energy, $\Sigma_{\rm sc}^{(b:1)}$. 
Since the excitron's polarization fluctuations are one-dimensional, 
the superconducting fluctuations are likewise one-dimensional.
}
\label{fig:05}
\end{figure*}

The contribution $\Sigma_{\rm sc}^{(b)}(K)$ 
in Fig.~\ref{fig:02}(b) is
\begin{align}
\label{eq:B1}
\Sigma_{\rm sc}^{(b)}(K)\!=\!-\sum_Q \sum_{K'} \sum_{K''}& 
\Gamma_{\rm ex}^{\uparrow \downarrow}(K,K'\!:\!Q\!-\!K\!-\!K')G(K') G(-K'\!+\!Q) 
&\nonumber \\ & \times 
G(K'')G(-K''\!+\!Q)
\Gamma_{\rm ex}^{\uparrow \downarrow}
(K'',K\!:\!Q\!-\!K\!-\!K'')G(-K\!+\!Q)D_{\rm sc}(Q),
\end{align}
where $\Gamma_{\rm ex}^{\uparrow \sigma}(K,K'\!:\!Q)$ is the excitron forming 
interaction vertex obtained from the BSE with the electron-hole 
irreducible interaction $\tilde{I}$ [Fig.~\ref{fig:05}(a)]. With this vertex, 
the excitron self-energy $\Sigma_{\rm ex}(K)$ is
\begin{align}
\label{eq:B2}
\Sigma_{\rm ex}(K)\!=\!-\sum_Q \sum_{K'\sigma}
\Gamma_{\rm ex}^{\uparrow \sigma}(K,K'\!:\!Q)G(K')G(K'\!+\!Q) 
G(K\!+\!Q)V({\bm q}),
\end{align}
as shown in Fig.~\ref{fig:05}(b).

As detailed in Ref.~\cite{YT_2024}, phase-space restrictions render the excitron 
in the 3DHEG effectively one-dimensional. For convenience, we denote by $K_{\rm F} 
\equiv \{{\bm k},i\omega_n\}$ a fermionic state on the Fermi surface 
($|{\bm k}|=k_{\rm F}$, $\omega_n \to 0$). When $K \approx K_{\rm F}$  
and $|{\bm q}|\! \approx \!2k_{\rm F}$, the states $K+Q$, $K'$, and $K'\!+\!Q$ 
are all forced to lie near the Fermi surface, aligning their momenta longitudinally. 
This constraint fixes the kinematics to 
$Q\!=\!-2K$, $K'\!=\!K\!+\!\delta Q$, and $K'\!+\!Q\!=\!-K\!+\!\delta Q$, 
with $\delta Q$ a small longitudinal deviation. Then, the dominant lowest-order 
process for $\Sigma_{\rm ex}(K)$ is shown in Fig.~\ref{fig:05}(c) as 
$\Sigma_{\rm ex}^{(1)}(K)$. Here, only the down-spin polarization is considered 
for the up-spin self-energy, because antiparallel-spin scatterings dominate over 
parallel-spin ones in short-range physics~\cite{YT_1987}. 

From the structure of $\Sigma_{\rm ex}^{(1)}(K)$, one readily anticipates 
Cooper-pair formation between the up-spin incident electron 
and the down-spin electron in the polarization field, 
giving rise to $\Sigma_{\rm sc}^{(b:1)}(K)$, the lowest-order term 
in $\Sigma_{\rm sc}^{(b)}(K)$. 
This observation shows that the fluctuation wave vector $Q$ in 
$\Sigma_{\rm sc}^{(b)}(K)$ is in fact the one-dimensional $\delta Q$, 
as it is directly tied to the internal kinematics of excitron formation. 
Physically, this reflects the fact that transient superconducting motion 
can occur only along the pre-existing polarization field.

Under the same approximations used in Appendix A, $\Sigma_{\rm sc}^{(b)}(K)$ 
at $k\!=\!k_{\rm F}$ is reduced to
\begin{align}
\label{eq:B3}
\Sigma_{\rm sc}^{(b)}(k_{\rm F},i\omega_n)=-\frac{1}{2}i\tilde{\omega}_n 
A_{1}(k_{\rm F},i\omega_n),
\end{align}
with 
\begin{align}
\label{eq:B4}
A_{1}(k_{\rm F},i\omega_n)=\left [ \frac{T_c
\Gamma_{\rm ex}^{\uparrow \downarrow}(K_{\rm F},K_{\rm F}\!:\!-2K_{\rm F})}
{\tilde{\omega}_n^2}
\right ]^2\!\frac{T_c}{N_0AD^2}
\sum_{{\bm q}}\frac{1}{{\bm q}_{\parallel}^2\!+\!\epsilon/A}
\frac{1}{[{\bm q}_{\parallel}^2\!+\!\epsilon/(2A)\!+\!|\omega_n|/D]^2},
\end{align}
where only ${\bm q}_{\parallel}$, the longitudinal component of ${\bm q}$, appears. 

Using the same $a$ as in Eq.~(\ref{eq:A6}), Eq.~(\ref{eq:B4}) reduces to
\begin{align}
\label{eq:B5}
A_{1}(k_{\rm F},i\omega_n)\!=\!
\left [ \frac{T_c
\Gamma_{\rm ex}^{\uparrow \downarrow}(K_{\rm F},K_{\rm F}\!:\!-2K_{\rm F})}
{\tilde{\omega}_n^2}
\right ]^2\!
\frac{1}{N_0}
\left ( \frac{T_c}{\pi} \right )^2
\left [ \frac{\xi(T)}{D} \right ]^3
\frac{2\pi \xi(T)^2}{S_{\rm eff}}
\frac{2a+1}{a^3(a+1)^2},
\end{align}
where $S_{\rm eff}$ is the effective polarization area per excitron.
In the large $\omega_n$ limit, we obtain
\begin{align}
\label{eq:B6}
A_{1}(k_{\rm F},i\omega_n)\approx
\left [T_c
\Gamma_{\rm ex}^{\uparrow \downarrow}(K_{\rm F},K_{\rm F}\!:\!-2K_{\rm F})
\right ]^2(4\pi/S_{\rm eff})
(1/N_0)(T_c/\pi)^2\,\xi(T)\,\omega_n^{-6}/D.
\end{align}

\setcounter{equation}{0}
\renewcommand{\theequation}{D\arabic{equation}}
\section{Determination of $A_1(k_{\rm F},0)$ and the quasiparticle lifetime}

In the limit $\omega_n \to 0$, the quantities 
$\tilde{\omega}_n$ and $a$ approach $1/(2\tau)$ and $1/\sqrt{2}$, 
respectively, so that $A_{1}(k_{\rm F},i\omega_n)$ in Eq.~(\ref{eq:B5}) 
converges to a $T$-dependent value $A_{1}(k_{\rm F},0)$. 

Higher-order contributions to $\Sigma(K)$ arising from $D_{\rm sc}(Q)$ 
are expected to become increasingly important at low $\omega_n$. 
We therefore assume that Eq.~(\ref{eq:04}) remains applicable 
to $Z_{\rm sc}(k_{\rm F},i\omega_n)$, with the understanding 
that such higher-order effects are absorbed into $A_1(k_F,i\omega_n)$. 
We further assume that the numerical data for 
$Z_{\rm sc}(k_{\rm F},i\omega_n)$ over the entire $\omega_n$ range 
in Fig.~\ref{fig:01}(e) contain the full information needed 
to extract these contributions.

To determine $\tau$ and $A_1(k_F,0)$, we apply the following 
extrapolation procedure. For fixed $r_s$ and $T$, and under the assumptions 
described above, we define 
\begin{align}
\label{eq:C1}
F(\omega_n)\equiv \omega_n\,Z_{\rm sc}(k_{\rm F},i\omega_n)
=\frac{1}{2}\left (\omega_n+\frac{1}{2\tau} \right ) A_1(k_F,i\omega_n),
\end{align}
at discrete positive Matsubara frequencies. We then employ 
Pad\'{e} approximants to analytically continue $F(\omega)$ 
to the positive real axis and slightly into the negative-$\omega$ region, 
enabling us to locate the first zero, $-\omega_0$ ($\omega_0>0$), 
of $F(\omega)$ near the origin. Once $\omega_0$ is obtained, $\tau$ 
follows from $\tau^{-1}=2\omega_0$, and $A_1(k_{\rm F},0)$ 
is subsequently determined via $A_1(k_{\rm F},0)=2F(0)/\omega_0$. 



\end{document}